\documentstyle[11pt,newpasp,twoside,epsf]{article}

\begin{document}

\title{A High Fraction of Mergers in the cluster MS1054--03 at $z=0.83$}
\author{M. Franx, P. G. van Dokkum}
\affil{Leiden Observatory, PO Box 9513, 2300 RA Leiden, Netherlands}
\author{D. G. Fabricant}
\affil{Center for Astrophysics, 60 Garden Street, Cambridge, MA 02138, USA}
\author{ D. Kelson}
\affil
{DTM, Carnegie Institution of
Washington, 5241 Broad Branch Road, NW, Washington D.C., 20015}
\author{ G. D. Illingworth}
\affil{University of California Observatories,
University of California, Santa Cruz, CA 95064}

\def\hmin{h$^{-1}$}
\begin{abstract}

We used the {\sl Hubble Space Telescope} to take a wide field,
multicolor image  of the high redshift cluster MS1054--03
at $z=0.83$.  
The Keck telescope was used to take
200 redshifts in the field. We have a total of 81 spectroscopically
confirmed cluster members with deep WFPC2 imaging.  A surprisingly
large number of these cluster members are mergers: 13 out of 77
galaxies are classified as merger remnants, or mergers in progress.
Most of these are red, and massive. Their restframe B
luminosities are amongst the brightest in the cluster: we find that 5
out of the brightest 16 are mergers.  No such bright mergers were
found in a lower redshift cluster with similar wide field coverage.
The mergers in MS1054--03 are preferentially found in the outskirts
of the cluster.
The distribution of pairs in the outskirts of the cluster shows an
excess of galaxies with separation $< 10$ \hmin kpc, which is
independent confirmation of the enhanced interaction rate.
The fraction of mergers (17 \%) is comparable to the fraction of
ellipticals (22 \%). 
Hence up to 50 \% of low redshift ellipticals may have formed through
mergers since $z=1$.
There is no evidence for strong
star formation in most of the mergers. The mean stellar ages of the mergers
may therefore be much older than the ``assembly age'', i.e., the 
age since the time the system was put together. These new results
are qualitatively consistent with predictions from hierarchical galaxy
formation, and are inconsistent with an early collapse for all
early type galaxies.
\end{abstract}


\section{Introduction}
High redshift clusters can be used very efficiently to study galaxy
formation and evolution.
Their high overdensities allow us to study large numbers of galaxies
with relatively small field imagers and spectrographs. Furthermore,
the range in morphologies in clusters, and between clusters, allow
us to study a variety of galaxy types.

We have embarked on a study of several massive clusters, out to a
redshift of 1, using the {\sl Hubble Space Telescope} to take mosaics
of deep multi-color images, and using large telescopes to take deep
spectroscopy. Our goal is to study a few clusters very well, with
wide coverage HST imaging, and extensive spectroscopy. Our study is
complementary to most other programs which study larger samples of
clusters with limited coverage (e.g., Dressler et al. 1997, and
references therein).

Until now, we have obtained wide field HST data on MS1358+62 at
$z=0.33$, MS1054--03 at $z=0.83$, and recently, MS2053--04 at
$z=0.58$. All these clusters were selected from the EMSS 
survey (Gioia et al. 1990). We have obtained $> 200$ redshifts
for each of these fields (Fisher et al. 1998, van Dokkum et al. 1999,
2000), and deep spectroscopy of the brighter 
galaxies to measure internal velocity dispersions (van Dokkum \&
Franx 1996, Kelson et al. 1999,
van Dokkum et al. 1998).
Here we present our new results on MS1054--03, the highest redshift
cluster in the EMSS catalogue (Gioia et al. 1995, Luppino \& Gioia 1996).

\section{Observations of MS1054--03}
We have taken deep, multicolor images of MS1054--03 at 6 pointings
with WFPC2 on the {\sl Hubble Space Telescope}.
The Keck telescope was used to take 200 spectra, aimed to be complete
to an  I magnitude of 22.7. The typical  integration time per galaxy was
40 min. We were able to measure redshifts of 186
galaxies, and of those, 80 were cluster members. Together with data
from literature, we found 89 cluster members, of which 81 lie in the
area of the HST images.

\subsection{Merger fraction}

We classified the spectroscopically confirmed cluster members,
analogous to our classification of galaxies in MS1358+62 (Fabricant et
al. 1999). We classified  galaxies along the revised Hubble sequence.
We allowed for a separate catagory of mergers.
We combined the 3 classifications from 3 of us, and verified that the
results were robust from classifier to classifier.
The results have been presented in van Dokkum et al. (1999, 2000).

The main outcome of this exercise is the high fraction of mergers
in MS1054--03. Many of these mergers are very luminous. One of the
most striking ways to display the effect, is to show a panel with
the 16 brightest galaxies (Fig 1.) Five out of these 16 were classified
as mergers.
A color version of figures 1 and 2 can be found at 
http://www.strw.leidenuniv.nl/\~\ franx/clusters.

\begin{figure}[t]
\vskip 6truecm
Here goes ms1054.bright16.gif. \hfill\break
See
http://www.strw.leidenuniv.nl/\~\ franx/clusters/ms1054
for the postscript file.
\vskip 5truecm
\caption{Brightest 16 galaxies in  MS1054--03 at $z=0.83$. These galaxies are
spectroscopically confirmed cluster members, and selected on the basis
of their restframe B luminosity. Five  out of 16 are classified as mergers.
The fraction of mergers remains similar if galaxies are corrected for
the luminosity brightening.
Each image has a physical
scale 30 \hmin kpc on a side.
}
\end{figure}
\begin{figure}[t]
\vskip 6truecm
Here goes ms1358.bright16.gif. \hfill\break
See
http://www.strw.leidenuniv.nl/\~\ franx/clusters/ms1054
for the postscript file.  
\vskip 5truecm
\caption{Brightest 16 galaxies in MS1358+62 at $z=0.33$, selected in
the same way as the galaxies in MS1054--03. Each image has a physical
scale of 30 \hmin kpc on a side.}
\end{figure}

A similar mosaic of the cluster MS1358+62 at $z=0.33$ is shown in
Fig. 2. 
The absence of mergers in this lower redshift cluster, and the much
more homogeneous color distribution are notable.

The enhancement of peculiar galaxies in MS1054--03 could be due
to the brightening of low mass galaxies during a starburst.
 We verified that the merger fraction
remains high if the galaxies are selected by mass, instead of (blue)
luminosity.

The optical colors of the mergers are consistent with this result.
As shown in Fig. 3, the mergers are generally red, with a few
exceptions. Similarly, the spectra of most of the mergers do not show
strong emission lines.
\begin{figure}
\plottwo{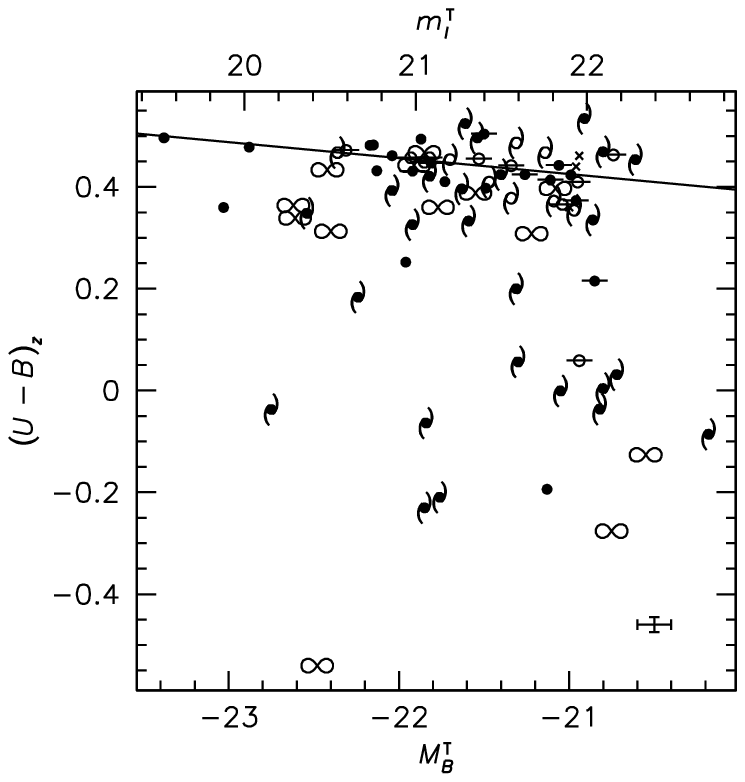}{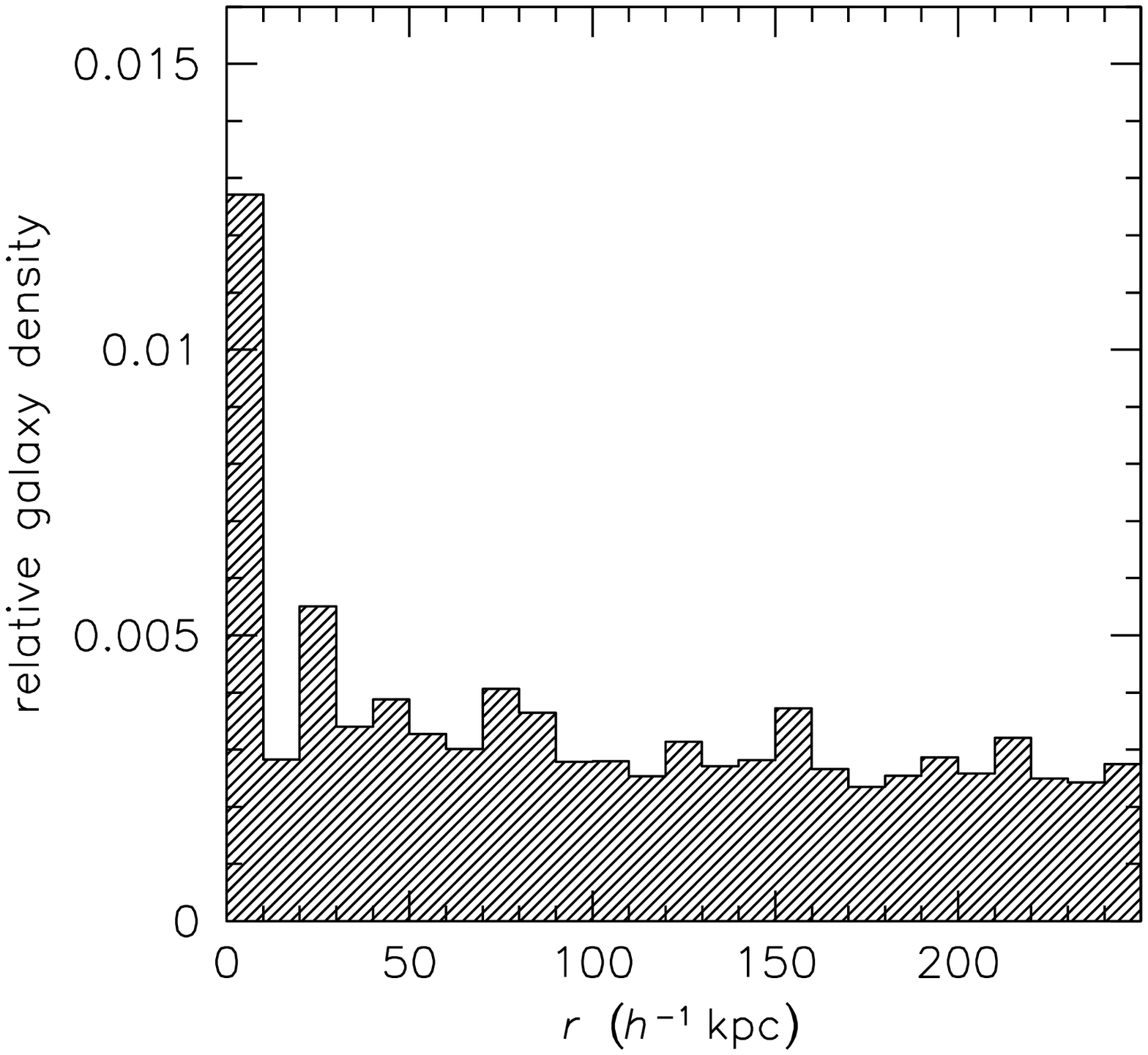}
\caption{(a) The color magnitude diagram for confirmed cluster members
of MS1054--03. The mergers are indicated with $\infty$ symbols.
They are generally red, although slightly bluer than ellipticals
and S0's. After aging to $z=0$, the scatter in the sample of
ellipticals + mergers will be identical to that of ellipticals in
nearby clusters.\hfill\break
(b) The overdensity  of pairs in the outskirts 
of MS1054--03. 
There is a clear excess of pairs at small separations $< 10$ \hmin
kpc.
This is independent confirmation of the enhanced interaction rate in
MS1054--03.}
\end{figure}

These results suggest that the bulk of the stars of the mergers were
formed well before the merger. Hence the stellar age of the merged
galaxies is significantly different from the ``assembly age'', i.e.,
the time at which the galaxy ``was put together''.

The results are consistent with the hypothesis that the mergers evolve
into ellipticals. Their scatter in the color-magnitude diagram is
significantly larger than the scatter for the ellipticals (0.073
versus 0.045 in restframe U-B). After aging of the stellar
populations, the scatter of the total population of
mergers+ellipticals will have decreased from 0.054 at $z=0.83$ to
0.015 at $z=0$.  Hence a low scatter at $z=0$ does not mean
that all galaxies in the population are homogeneous and very old: the
influence of merging can be small if the star formation involved with
the merging is low.

The physical  reason for the low star formation is unknown: it is
possible that the massive precursor galaxies had already lost their
cold gas due to internal processes (such as super winds, or
winds driven by nuclear activity). Alternatively, the cluster
environment
may play an important role: the cold gas may have been
stripped
by the cluster X-ray gas.
Observations of more clusters may shed further light on this effect.

\subsection{Pair fraction}

Whereas the classifications of galaxies remains a subjective
procedure, counting close pairs of galaxies is a very objective way
to establish whether interactions and mergers are
enhanced. Furthermore,
the distribution of pairs may shed light on the future merging rate in
the
cluster. 
We have counted the number of pairs in the outskirts of the cluster,
to avoid the high density core.
The pair fraction is shown in Fig 3b. As we can see, there is
an  excess of pairs at small separations ($<$ 10 \hmin kpc).
Half of these  are classified as mergers, the other half not.
These may constitute a reservoir of ``future'' mergers.
It will be interesting to measure the velocity differences of the
galaxies in pairs.

\section{Conclusions}
We have found a high fraction of mergers, which are generally red.
The fraction is comparable to the number of ellipticals.
The results are qualitatively consistent with
 the hypothesis of hierarchical galaxy
formation. The relatively old stellar age of the mergers compared to
the young ``assembly age'' is consistent with predictions from
semi-analytical models (e.g., Kauffmann 1996).

The results are inconsistent with the hypothesis that all ellipticals
are formed and assembled at very high redshift. Nevertheless, many
questions remain open: 
\begin{itemize}
\item
Is the result for MS1054--03 typical for high
redshift clusters ? Is the merger fraction higher or lower in the
field ?
It is interesting to note that studies of the field
give high merger fractions and/or pair fractions at intermediate
redshift (e.g., Patton et
al. 1995, Le Fevre et al. 1999). It remains to be seen whether these
field mergers are as massive as the mergers found in MS1054--03.

\item What is the
 typical redshift  at which the mass of early type galaxies
was half of the current mass ? 

\item When did the major episode of
star formation occur ?
\end{itemize}
Future studies can be directed to shed light
on these questions.


\end{document}